\documentclass{PoS}
\usepackage{graphicx}
\usepackage[dvips]{epsfig}
\usepackage{amssymb,amsmath}
\usepackage{slashed}
\usepackage{url}

\title{CERN-INO magical Beta-beam experiment: \\ A high precision probe 
for neutrino parameters}

\ShortTitle{CERN-INO Magical Beta-Beam}

 \author{\speaker{Sanjib Kumar Agarwalla}\thanks{It is my pleasure 
  to acknowledge the support of Harish-Chandra Research Institute (HRI) 
  where the work has been done. The computational work presented in this 
  talk has been performed using the HRI cluster facilities. 
  I would like to thank Anselmo Cervera for providing me the local 
  hospitality during Nufact08.} \\
         Department of Physics, Virginia Tech, \\
	 Blacksburg, VA 24061, USA \\
         E-mail: \email{sanjib@vt.edu}}

 \author{Sandhya Choubey  \\
         Harish-Chandra Research Institute, \\
	 Chhatnag Road, Jhunsi, Allahabad  211019, India \\
         E-mail: \email{sandhya@hri.res.in}}

 \author{Amitava Raychaudhuri  \\
         Harish-Chandra Research Institute, \\
	 Chhatnag Road, Jhunsi, Allahabad  211019, India \\
         E-mail: \email{raychaud@hri.res.in}}

\abstract{
This talk is an attempt to underscore in detail the physics reach
of an experimental set-up where neutrinos produced in a beta-beam 
facility at CERN would be observed in the proposed large magnetized 
iron calorimeter detector (ICAL) at the India-based Neutrino Observatory 
(INO). The ``magical'' CERN-INO beta-beam set-up offers an excellent 
avenue to use the ``Golden'' channel ($\nu_e \rightarrow \nu_{\mu}$) 
oscillation probability for a simultaneous determination of the neutrino 
mass ordering and $\theta_{13}$ avoiding the impact of the CP phase
$\delta_{CP}$ on these measurements. With Lorentz boost $\gamma=650$ and
irrespective of the true value of $\delta_{CP}$,
the neutrino mass hierarchy could be determined at $3\sigma$ C.L.
if $\sin^22\theta_{13}{\rm {(true)}} > 5.6 \times 10^{-4}$
and we can expect an unambiguous signal for $\theta_{13}$
at $3\sigma$ C.L. if $\sin^22\theta_{13}{\rm {(true)}} > 5.1 \times 10^{-4}$
independent of the true neutrino mass hierarchy.} 

\FullConference{10th International Workshop on Neutrino Factories, Super beams and Beta beams\\
		 June 30 - July 5 2008\\
		 Valencia, Spain}

\newcommand{\ma}{\Delta m^2_{31}}
\newcommand{\stch}{\sin^2 2\theta_{13}}
\newcommand{\stcht}{\sin^2 2\theta_{13}{\mbox {(true)}}}
\newcommand{\dcpt}{\delta_{CP}{\mbox {(true)}}}
\def\nue{{\nu_e}}
\def\anue{{\bar\nu_e}}
\def\numu{{\nu_{\mu}}}
\def\anumu{{\bar\nu_{\mu}}}

\begin{document}

\section{Introduction}

Future long baseline neutrino oscillation experiments will play a crucial
role to measure the third mixing angle $\theta_{13}$, the 
sign$\footnote{The neutrino mass hierarchy is termed ``normal 
(NH)'' (``inverted (IH)'') if $\ma = m_3^2 - m_1^2$ is positive (negative).}$ 
of $\Delta m^2_{31}\equiv m_3^2 - m_1^2$ ($sgn(\ma)$) and the CP phase 
($\delta_{CP}$) using the $\nue \to \numu$ transition probability
($P_{e\mu}$), often referred to in the literature as the 
``golden channel'' \cite{golden}. An absolutely pure and intense 
$\nue$ (or $\anue$) flux can be produced using ``beta-beams'' \cite{zucc} 
at CERN and oscillations can be observed through muons produced {\it{via}} 
$\numu$ (or $\anumu$) at ICAL@INO \cite{ino}. Interestingly, the CERN-INO 
distance of 7152 km happens to be tantalizingly close to the so-called 
``magic baseline'' \cite{eight,magic} where the sensitivity to the neutrino 
mass ordering ($sgn(\ma)$) and more importantly, $\theta_{13}$, goes up 
significantly, while the sensitivity to the unknown CP phase is absent. 
This permits such an experiment involving the golden $P_{e\mu}$ channel to make 
precise measurements of the mixing angle $\theta_{13}$ and neutrino mass hierarchy 
avoiding the issues of intrinsic degeneracies \cite{degeneracy} and correlations \cite{golden}
which plague other baselines. This large baseline also captures a near-maximal matter-induced 
contribution to the oscillation probability. In this talk, we will
discuss in detail the physics prospects of this CERN-INO magical beta-beam project.  

\section{``Golden channel'' oscillations}

The expression for $P_{e\mu}$ in matter, upto second order terms in the 
small quantities $\alpha \equiv \Delta m_{21}^2/\Delta m_{31}^2$ and $\theta_{13}$,
is given by \cite{golden,freund}:
{\footnotesize{
\begin{eqnarray}
  P_{e\mu} & \simeq & \sin^2 2\theta_{13} \, \sin^2 \theta_{23}
  \frac{\sin^2[(1- \hat{A}){\Delta}]}{(1-\hat{A})^2}
  \pm \alpha  \sin 2\theta_{13} \,  \xi \sin \delta_{CP}
  \sin({\Delta})  \frac{\sin(\hat{A}{\Delta})}{\hat{A}}
  \frac{\sin[(1-\hat{A}){\Delta}]}{(1-\hat{A})}
  \nonumber \\
  &+& \alpha  \sin 2\theta_{13} \,  \xi \cos \delta_{CP}
  \cos({\Delta})  \frac{\sin(\hat{A}{\Delta})}{\hat{A}}
  \frac{\sin[(1-\hat{A}){\Delta}]} {(1-\hat{A})}
  + \alpha^2 \, \cos^2 \theta_{23}  \sin^2 2\theta_{12}
  \frac{\sin^2(\hat{A}{\Delta})}{\hat{A}^2},
\label{equ:pemu}
\end{eqnarray}
}}
where $\Delta \equiv \Delta m_{31}^2 L/(4 E)$, $\xi \equiv \cos\theta_{13} \,
\sin 2\theta_{12} \, \sin 2\theta_{23}$,  and $\hat{A} \equiv \pm (2 \sqrt{2}
G_F N_e E)/\Delta m_{31}^2$.  $G_F$ and $N_e$ are the Fermi coupling constant
and the electron density in matter, respectively. The second term of 
Eq. \ref{equ:pemu} is positive (negative) for neutrinos (antineutrinos).
The sign of $\hat{A}$ is positive (negative) for neutrinos (antineutrinos) 
with NH and it is opposite for IH.

\section{The CERN-INO magical Beta-beam set-up}

A particularly interesting scenario arises when $\sin(\hat{A}\Delta)=0$, 
the last three terms in Eq. \ref{equ:pemu} drop out and the $\delta_{CP}$ 
dependence disappears from the $P_{e\mu}$ channel which provides a clean 
ground for the determination of $\theta_{13}$ and $sgn(\ma)$. 
Since $\hat{A}\Delta = \pm  (2 \sqrt{2} G_F N_e L)/4$ by definition, 
the first non-trivial solution for $\sin(\hat{A}\Delta)=0$ reduces 
to $\rho L = \sqrt{2}\pi/G_F Y_e$, where $Y_e$ is the electron fraction 
inside the Earth. This gives $\frac{\rho}{[{\rm g/cc}]}\frac{L}{[km]} 
\simeq 32725~$, which for the PREM \cite{prem} density profile of the earth
is satisfied for the ``magic baseline'', $L_{\rm magic} \simeq 7690 ~{\rm km}$.
The CERN-INO distance corresponds to $L=7152$ km, which is pretty close to 
this magic baseline. 

\begin{figure}[!t]
\includegraphics[width=7.25cm,height=.28\textheight]{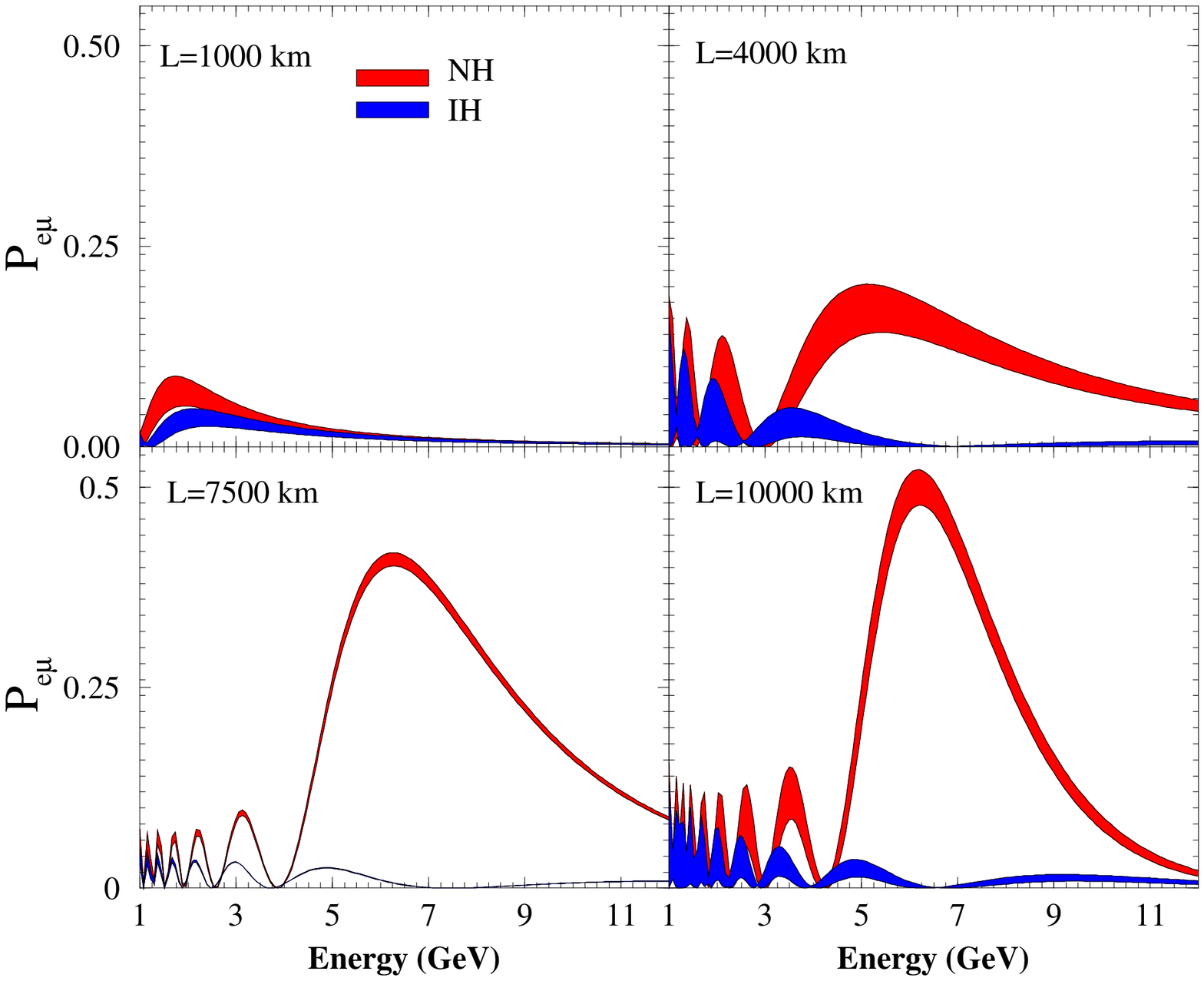}
\hglue 0.5cm
\includegraphics[width=7.25cm,height=.28\textheight]{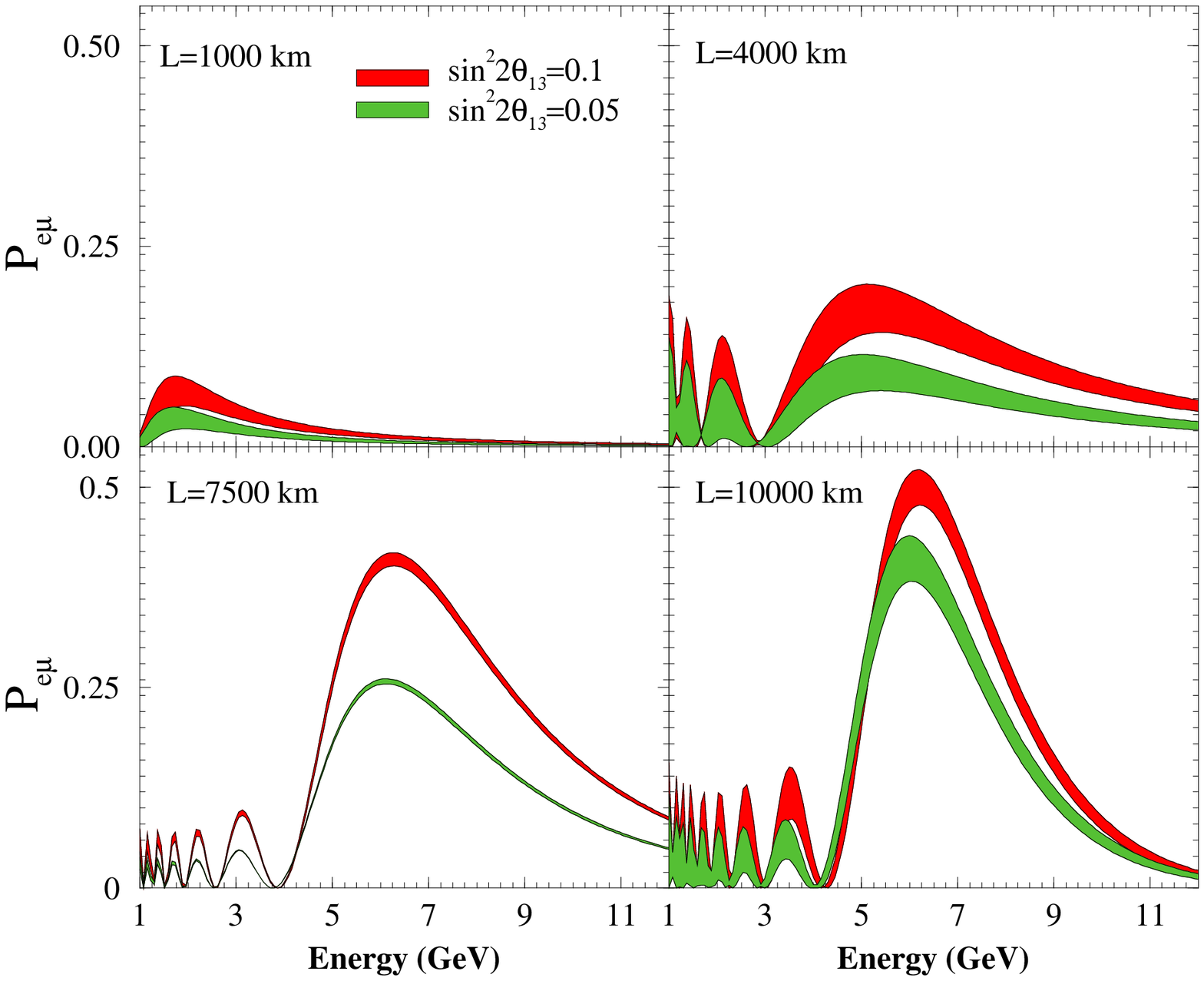}
\caption{\label{fig:prob}
Both the panels show the energy dependence of $P_{e\mu}$
for four baselines where the band reflects the effect of the
unknown $\delta_{CP}$. Left panel depicts the effect of $\delta_{CP}$
in making distinction between NH $\&$ IH with $\stch=0.1$.
Right panel reflects the difference in $P_{e\mu}$ for two different values
of $\stch$ with NH.}
\end{figure}

\begin{figure}[!t]

\includegraphics[width=7.25cm,height=.28\textheight]{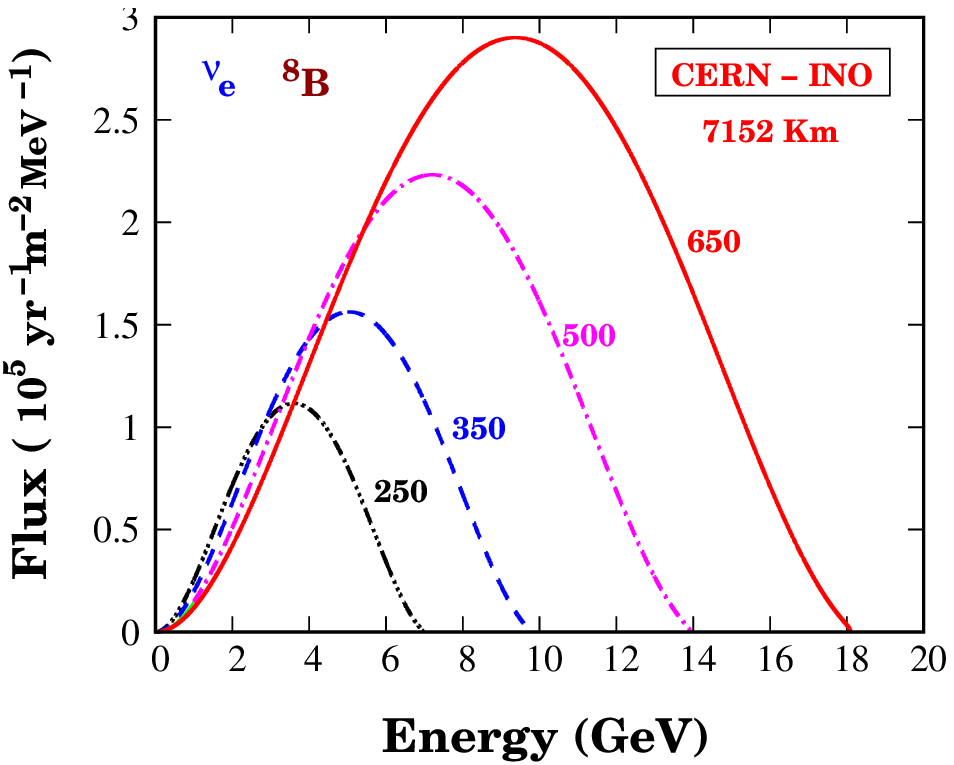}
\hglue 0.5cm
\includegraphics[width=7.25cm,height=.28\textheight]{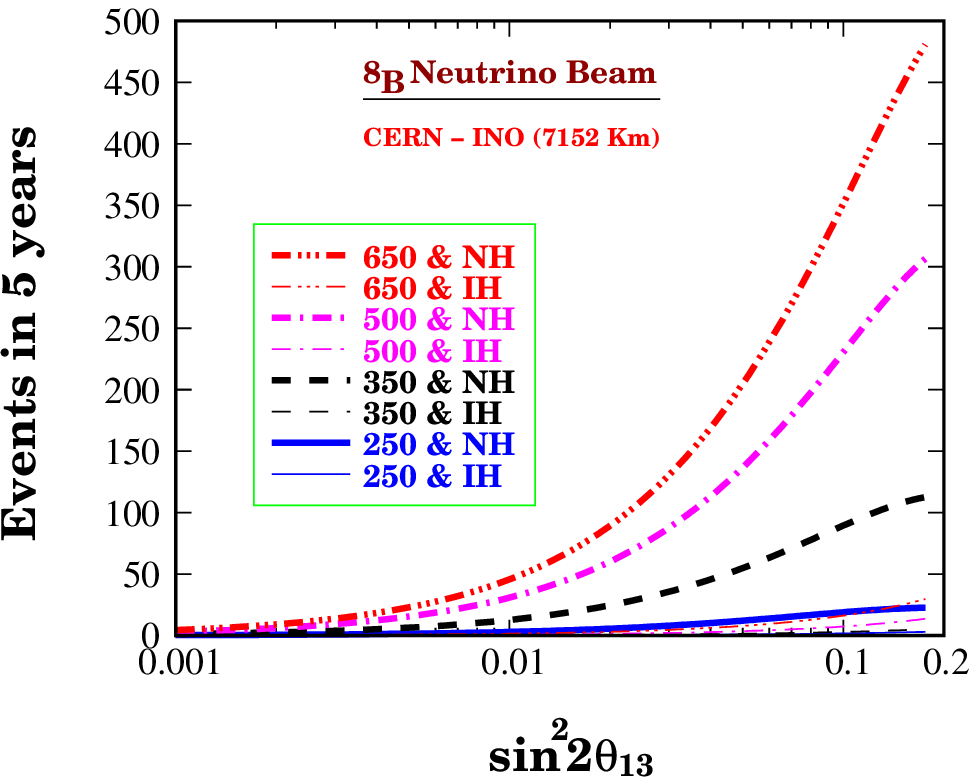}
\caption{\label{fig:flux_rate}
Left panel shows the boosted unoscillated spectrum of neutrinos
from $^8$B ion which will hit the ICAL@INO detector, for four different
benchmark values of $\gamma$. Right panel depicts the expected 
$\mu^{-}$ events in 5 years as a function of $\stch$ 
with a total detector mass of 50 kton and 80\% detection efficiency.
The value of $\gamma$ and the hierarchy chosen corresponding to each
curve is shown in the figure legend.}
\end{figure}

This large baseline requires traversal through denser 
regions of the earth. Thus, for neutrinos (antineutrinos) with energies in the
range 3-8 GeV sizable matter effects are induced for NH (IH). A unique aspect of 
this set-up is the possibility of observing near-resonant matter effects in the 
$P_{e\mu}$ channel. In fact, for this baseline, the average Earth matter 
density calculated using the PREM profile is $\rho_{av}=4.17$ g/cc, for which
the resonance energy is 
$E_{res} \equiv |\ma| \cos 2\theta_{13} / 2\sqrt{2} G_F N_e
= 7.45$ GeV, taking $|\ma|=2.5\times 10^{-3}$ eV$^2$ and $\stch=0.1$.

We present all our results assuming certain benchmark values 
\cite{limits} for the oscillation parameters 
$|\Delta m^2_{31}| = 2.5 \times 10^{-3} \ {\rm eV}^2$,
$\sin^2 2 \theta_{23} = 1.0$,
$\Delta m^2_{21} = 8.0 \times 10^{-5} \ {\rm eV}^2$,
$\sin^2\theta_{12} = 0.31$ and $\delta_{CP} = 0$.
The exact neutrino transition probability $P_{e\mu}$ using the PREM density
profile is given in Fig. \ref{fig:prob}. For neutrinos (antineutrinos),
matter effects in long baselines cause a significant enhancement 
in $P_{e\mu}$ for NH (IH), while for IH (NH), the probability is almost unaffected.
This feature can be used to determine the neutrino mass hierarchy 
(see left panel of Fig. \ref{fig:prob}). For $L=7500$ km, which is close to the 
magic baseline, the effect of the CP phase is seen to be almost negligible.
This allows a clean measurement of $sgn(\ma)$ (see left panel of Fig. \ref{fig:prob})
and $\theta_{13}$ (see right panel of Fig. \ref{fig:prob}), while at all other baselines 
the impact of $\delta_{CP}$ on $P_{e\mu}$ is appreciable.

Zucchelli \cite{zucc} put forward the novel idea of a beta-beam 
\cite{betapemu,betaino1,betaino2,betaino3,rparity1,rparity2,pee,bboptim,twobaseline,winter}, 
which is based on the concept of creating a pure, well understood, intense,
collimated beam of $\nue$ or $\anue$ through the beta-decay of completely
ionized radioactive ions. We consider $^8$B ($^8$Li) ion \cite{rubbia}
as a possible source for a $\nue$ ($\anue$) beta-beam. The end point energies of
$^8$B and $^8$Li are $\sim$ 13-14 MeV. For the Lorentz boost factor $\gamma=250(500)$
the $^8$B and $^8$Li sources have peak energy around $\sim 4(7)$ GeV. 
We can see from Fig. \ref{fig:flux_rate} (left panel) that with $\gamma$ = 500, 
the $\nue$ spectrum peaks nearby $E_{res}$. We assume that it is possible to get
$2.9\times 10^{18}$ useful decays per year for $^8$Li and $1.1\times 10^{18}$ for 
$^8$B for all values of $\gamma$.

The ICAL@INO detector, capable of detecting muons along with their charge, 
is planned to have a total mass of 50 kton at startup, which might be later 
upgraded to 100 kton. The INO facility is expected to come 
up at PUSHEP (lat. North 11.5$^\circ$, long. East 76.6$^\circ$), situated close 
to Bangalore in southern India. This constitutes a baseline of 7152 km from CERN. 
The detector will be made of magnetized iron slabs with interleaved active detector 
elements. For ICAL, glass resistive plate chambers have been chosen as the active 
elements. The expected number of events are shown in the right panel of Fig. 
\ref{fig:flux_rate}. We take a detector with an energy threshold of 1 GeV,
detection efficiency of 80\% and charge identification efficiency of 95\%.

\begin{figure}[!t]
\includegraphics[width=7.25cm,height=.28\textheight]{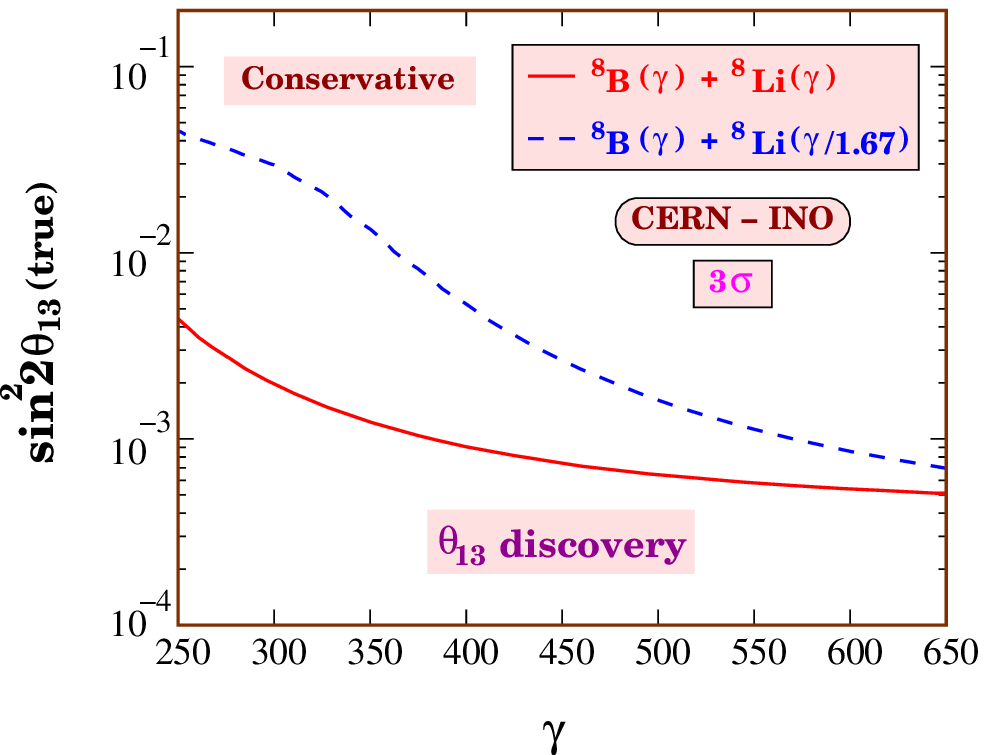}
\hglue 0.5cm
\includegraphics[width=7.25cm,height=.28\textheight]{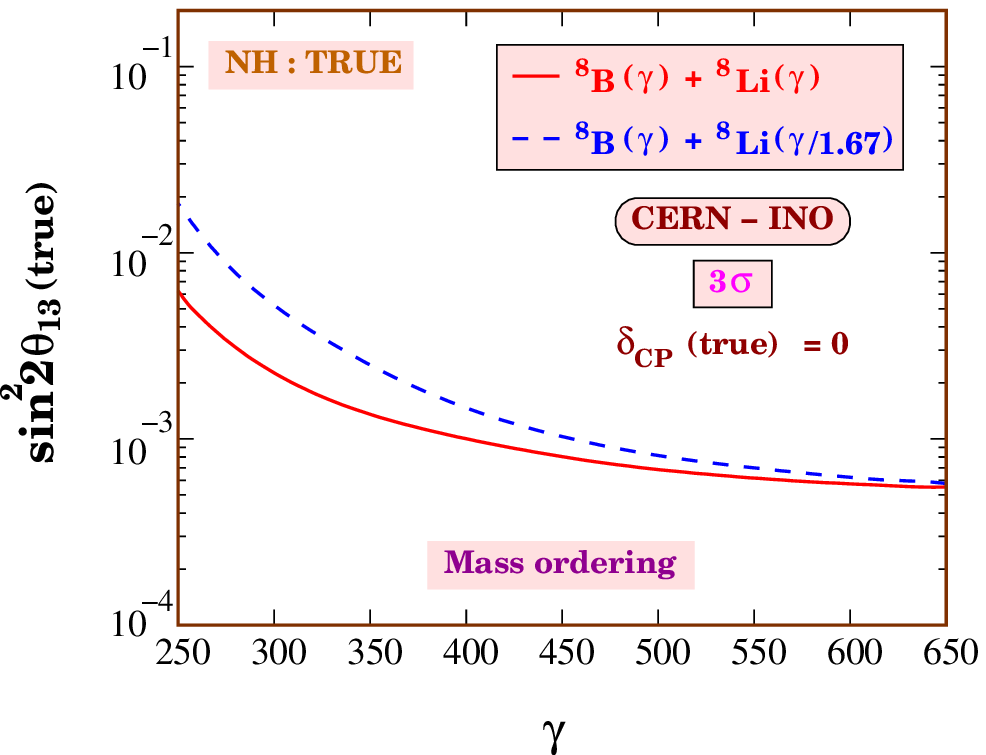}
\caption{\label{fig:senshier}
Left panel shows the $3\sigma$ discovery reach for $\stcht$. Right
panel shows the minimum value of $\stch$(true) for which the wrong
inverted hierarchy can be ruled out at the $3\sigma$ C.L., as a function of
the Lorentz boost $\gamma$. The red solid lines in both the panels are obtained
when the $\gamma$ is assumed to be the same for both the neutrino and
the antineutrino beams. The blue dashed lines
show the corresponding limits when the $\gamma$ for the
$^8$Li is scaled down by a factor of 1.67 with respect to the
$\gamma$ of the neutrino beam, which is plotted in the $x$-axis.
}
\end{figure}

The $\theta_{13}$ sensitivity reach is defined as the range of $\stch$
which is incompatible with the data generated for $\stcht=0$ at the
$3\sigma$ C.L. This performance indicator corresponds to the new $\stch$ 
limit if the experiment does not see a signal for $\theta_{13}$-driven oscillations.
At $3\sigma$, the CERN-INO beta-beam set-up can constrain
$\sin^22\theta_{13} < {1.14\times 10^{-3}}$ with five years of running of 
the beta-beam in both polarities with the same $\gamma=650$ and full spectral information. 
The $\theta_{13}$ discovery reach is defined as the range of $\stcht$ values 
which allow us to rule out $\stch=0$ at the $3\sigma$ C.L. We present our results 
in the left panel of Fig. \ref{fig:senshier}, as a function of $\gamma$. 
The plot presented show the most conservative numbers which have been obtained
by considering all values of $\dcpt$ and both hierarchies.
The mass hierarchy sensitivity is defined as the range of $\stcht$ for 
which the wrong hierarchy can be excluded at the $3\sigma$ C.L.
The results are depicted as a function of $\gamma$ in the right
panel of Fig. \ref{fig:senshier}. For NH true, the $sgn(\ma)$ reach corresponds to
$\stcht > {{5.51 \times 10^{-4}}}$, with 5 years energy binned data of 
both polarities and $\gamma=650$. Here we had assumed $\dcpt=0$. 
However, as discussed before, the effect of $\delta_{CP}$ is minimal due to
the near magic baseline and hence we expect this sensitivity 
to be almost independent of $\dcpt$.

\section{Conclusions}

In this talk, we discussed the expected physics performance of
the CERN-INO magical beta-beam set-up. The results are very 
impressive, comparable to those expected in neutrino factories.


\end{document}